\documentstyle[12pt]{article}
\begin{document}

\title{Bose-Hubbard Hamiltonian from generalized commutation rules }
\author{J. C. Flores}
\date{Universidad de Tarapac\'a \\
Departamento de F\'{\i}sica \\
Casilla 7-D, Arica \\
Chile}
\maketitle

\baselineskip=20pt

In a first order approximation, the Bose-Hubbard Hamiltonian with on-site
interaction is obtained from the free Hamiltonian ($U=0$) and generalized
commutation relations for the annihilation-creation operators. Similar
generalized commutation relations were used for the first time in high
energy physics. The spectrum of the system can be found formally by using
the algebraic properties of the generalized operators. 
$$
{} 
$$

$$
$$


\newpage

In recent years much attention has been given to the Hubbard model for
bosons ([1-5] and references therein). At first glance, this is a simple
descriptions for interacting particles, and related to a subject like phase
transitions. In this brief report we obtain, in a first order, the
Bose-Hubbard model from usual phonon-like Hamiltonian where the annihilation
and creation operators satisfy generalized canonical commutation relations.
In fact, the Bose-Hubbard model corresponds to the limit of small
perturbation with respect to usual commutation rules. To the best of our
knowledge this connection is not found in the literature. Curiously, the
model is formally solvable in a way similar to the free-phonon system. So,
this work opens new scopes about the interpretation of the Bose-Hubbard
Hamiltonian, and its relationship with some ideas in high energy physics.
The feeling of the report is that interaction terms can be considered
modifying the commutation rules for the relevant operators. 
$$
{} 
$$

Consider the Hamiltonian ${\cal H}$ and the operator ${\cal N}$ defined by
(we use units in which $\hbar =1$) 
\begin{equation}
{\cal H}=\frac 1L\sum_k\omega _k(\beta _k^{\dagger }\beta _k+\beta _k\beta
_k^{\dagger }), 
\end{equation}
\begin{equation}
{\cal N}=\frac 1L\sum_k(\beta _k^{\dagger }\beta _k+\beta _k\beta
_k^{\dagger }), 
\end{equation}
where we assume $\omega _k$ a real function of the wavenumber $k$ (first
Brillouin zone) and $L$ is the total number of sites in the lattice.
Moreover, consider the operators $\beta _k$ and $\beta _k^{\dagger }{}$
satisfying the implicit commutation relations 
\begin{equation}
[\beta _k,\beta _p^{\dagger }]=(L+\frac U{\omega _k}{\cal N})\delta _{k,p}. 
\end{equation}
where $\delta _{k.p}$ is the usual Kronecker symbol and $U$ is a constant
which will be related to the usual strength-interaction-parameter in the
Bose-Hubbard model. Moreover we assume

\begin{equation}
\begin{array}{c}
\lbrack \beta _k,\beta _p]=[\beta _k^{\dagger },\beta _p^{\dagger }]=0. \\  
\end{array}
\end{equation}

With respect to the above definitions we remark : (a) Similar
generalizations of the canonical commutation relations were proposed
originally in high energy physics to explain some phenomena related to quark
physics [6-8] where the correction was proportional to the Hamiltonian
rather than to the operator ${\cal N}$ . (b) The case $U=0$ corresponds to
the well-known model of free excitations with spectrum $\omega _k$, where $%
\beta _k^{\dagger }$ and $\beta _k$ are the usual bosonic operators. (c) We
can show $[{\cal H,N}]=0$ [9] and then, the commutation relations (3) do not
change with time. (d) If we consider the limit $L\rightarrow \infty $ then $%
\frac 1L\sum_k\rightarrow \int dk$ in (1) and (2). Moreover, in this limit
we have $L\delta _{k,p}\rightarrow \delta (k-p)$ and a correction
proportional to the density $\frac{{\cal N}}L$ in the commutator (3).%
$$
{} 
$$

To determine the first order relationship between the system defined by
(1-3) and the Bose-Hubbard model we go on as follows: let $b_k$ and $%
b_k^{\dagger }$ be the usual creation-annihilation Bose operators which
commutator $[b_k,b_p^{\dagger }]=L\delta _{k,p}$. Since $U\to 0$ we expect $%
\beta _k\to b_k$ ; then, it seems reasonable to assume the first-order
expansion in $U$, 
\begin{equation}
\beta _k=b_k+\frac U{2\omega _k}\sum_{prs}f(k,p,r,s)b_pb_r^{\dagger
}b_s+O(U^2)
\end{equation}
where $f$ is a unknown real function which will be determined using, at
first order, the generalized commutation rules (3) (i.e. we put (5) in (3)).
After some calculations we get 
\begin{equation}
\beta _k=b_k+\frac U{2\omega _kL^2}\sum_{prs}\delta
_{k+r,p+s}b_pb_r^{\dagger }b_s+O(U^2)
\end{equation}
Namely, the expansion (5) is consistent with the commutation rules (3). An
homogenous zero order groundstate was assumed. In this approximation (6),
the Hamiltonian (1) becomes

\begin{equation}
{\cal H}=\frac 1L\sum_k\omega _k(b_k^{\dagger }b_k+b_kb_k^{\dagger })+\frac
U{L^3}\sum_{krps}\delta _{k+r,p+s}\left( b_kb_s^{\dagger }b_rb_p^{\dagger
}+b_k^{\dagger }b_pb_r^{\dagger }b_s\right) +O(U^2). 
\end{equation}
And if we consider the Fourier transform to real space $l$ in the lattice

\begin{equation}
b_l=\frac 1L\sum_kb_ke^{ikl} 
\end{equation}
then, the Hamiltonian (7) can be written at first order in $U$ as 
\begin{equation}
{\cal H}=-\sum_lt(b_l^{\dagger }b_{l-1}+b_lb_{l+1}^{\dagger
}+b_lb_{l-1}^{\dagger }+b_l^{\dagger }b_{l+1})+U\sum_l(b_l^{\dagger
}b_lb_l^{\dagger }b_l+b_lb_l^{\dagger }b_lb_l^{\dagger })+O(U^2) 
\end{equation}
which corresponds to the symmetrized Bose-Hubbard Hamiltonian of strength $U$
in position representation. To obtain (9) the usual dispersion relation in
one-dimension $\omega _k=-2tcosk,$ where $t$ is the hopping parameter, was
assumed. It must be noted, however, that the Hubbard term is also obtained
for dimension greater than one ($D>1$).

$_{}$%
$$
{} 
$$

To determine the spectrum of the system (1-3) we consider the eigenvector of 
${\cal H}$ and ${\cal N}$ given by $|E_{k,n};N_{k,n}\rangle $. Namely,

\begin{equation}
{\cal H}\mid E_{k,n};N_{k,n}\rangle =E_{k,n}\mid E_{k,n};N_{k,n}\rangle , 
\end{equation}

\begin{equation}
{\cal N}\mid E_{k,n};N_{k,n}\rangle =N_{k,n}\mid E_{k,n};N_{k,n}\rangle . 
\end{equation}
To find the relationship between different eigenvalues of ${\cal H}$ (i.e.
between $E_{k,n+1}$ and $E_{k,n})$ we go on as follows : by using the
commutation rules (3), one finds 
\begin{equation}
{\cal H}\beta _k^{\dagger }|E_{k,n};N_{k,n}\rangle =\{E_{k,n}+2\omega
_k+\frac UL({\cal N}\beta _k^{\dagger }+\beta _k^{\dagger }{\cal N}%
)\}|E_{k,n};N_{k,n}\rangle . 
\end{equation}
However, also from equation (3) one has 
\begin{equation}
{\cal N}\beta _k^{\dagger }|E_{k,n};N_{k,n}\rangle =\frac{2+N_{k,n}(1+\frac
U{L\omega _k})}{1-\frac U{L\omega _k}}\beta _k^{\dagger
}|E_{k,n};N_{k,n}\rangle 
\end{equation}
then $\beta _k^{\dagger }\mid E_{k,n};N_{k,n}\rangle $ is an eigenvector of $%
{\cal N}$. Expressions (13) gives the relationship between different
eigenvalues $N_{k,n}$ of the operator ${\cal N}$ : 
\begin{equation}
N_{k,n+1}=\frac{2+N_{k,n}(1+\frac U{L\omega _k})}{1-\frac U{L\omega _k}}. 
\end{equation}
Introducing (13) onto (12), we have ${\cal H}\beta _k\mid
E_{k,n};N_{k,n}\rangle =E_{k,n+1}\beta _k\mid E_{k,n};N_{k,n}\rangle $ and
the relationship for the spectrum of ${\cal H}$ 
\begin{equation}
E_{k,n+1}=E_{k,n}+2\omega _k+\frac{2U}{L-\frac U{\omega _k}}(N_{k,n}+1) 
\end{equation}
where if $U=0$ becomes the usual oscillator-like-spectrum for phonons. The
spectrum (15) is not given in explicit form because of its dependence on the
eigenvalue $N_{k,n}$ . Nevertheless, from equation (14), it can be solved if
we note that 
\begin{equation}
N_{k,n}=\frac{L\omega _k}U\left[ \left( \frac{1+\frac U{L\omega _k}}{1-\frac
U{L\omega _k}}\right) ^n-1\right] 
\end{equation}
where it was assumed $N_{k,0}=0$.

$$
{} 
$$

We finally raise some questions related to the above connection between the
Bose-Hubbard Hamiltonian and the generalized commutation rules: (i) The
extension to the usual Fermi-Hubbard model was not worked-out. The
generalization of the commutation (or anticommutation) relations (3) is not
straightforward. (ii) Thermodynamic properties like internal energy, heat
capacity , thermal conduction, must be calculated by using the spectrum
(15). Nevertheless, this seems not direct, although one may attempt a
perturbation expansion in powers of $U$. (iii) The usual term related to
density-density nearest-neighbor interaction [2-5], between bosons, is
absent in (9). It seems that such a generalization requires a modification
of the commutation relations (3). (iv) In last years, much attention has
been given to disordered Hubbard models ([1, 10-13] and references therein),
in our case it is not clear how to consider disorder in the formulation
(1-3). Nevertheless, this can be partially carried-out using an appropriate
spectral function $\omega _k$ like to this one of disordered systems.

$$
{} 
$$

A more detailed treatment of points (i-iv) and further physical applications
will be given elsewhere. 
$$
$$

Acknowledgments : It is my pleasure to thanks useful discussions with C.
Elphick, I. Saavedra and S. Montecinos.

$$
{} 
$$


\begin{thebibliography}{99}
\bibitem{}  To see this we can write always ${\cal H}=\frac 1L\sum \omega
_k\eta _k$ and ${\cal N}=\frac 1L\sum \eta _k$ with $\eta _k=\beta _k\beta
_k^{\dagger }+\beta _k^{\dagger }\beta _k$. From (3) and (4) we have $[\eta
_k,\eta _p]=0$ for all $k,p$.
\end{thebibliography}
\end{document}